# Gas Mixture Diffusion and Distribution in the Porous ZIF-90 Framework

Ashok Yacham[†], Tarak K. Patra[†$*], Jithin John Varghese[†*]

[†]Department of Chemical Engineering, Indian Institute of Technology Madras, Chennai 600036, India
[$]Centre for Atomistic Modelling and Materials Design, Indian Institute of Technology Madras, Chennai 600036, India

**Abstract:**

Understanding how gas mixtures diffuse and distribute within porous frameworks is central to designing advanced separation and storage materials. Here, we investigate the transport and spatial distribution of binary gas mixtures in a porous metal organic framework, viz., ZIF-90, using molecular simulations. We perform grand canonical Monte Carlo (GCMC) simulations to examine the competitive adsorption of carbon dioxide ($CO_2$) and nitrogen ($N_2$) from a binary gas mixture in ZIF-90, while molecular dynamics (MD) simulations are conducted to investigate the transport behavior of the adsorbed molecules within the framework. These integrated simulations reveal that the framework topology and pore chemistry jointly dictate diffusion pathways and preferential occupancy of gas species, underscoring their intrinsic interdependence. Competitive adsorption leads to distinct spatial partitioning within the pores, which in turn modulates mixture diffusivity inside the porous medium compared to their bulk properties. Our results provide molecular-level insight into how ZIF-90 accommodates and separates gas mixtures, offering design principles for optimizing metal-organic frameworks in energy and environmental applications.

Keywords: Metal-Organic Framework, GCMC, MD, and Gas separation

**Table of Contents**

The present study offers fundamental insights into how the spatial distribution and diffusion dynamics of individual gas species are interrelated within the ZIF-90 metal–organic framework.

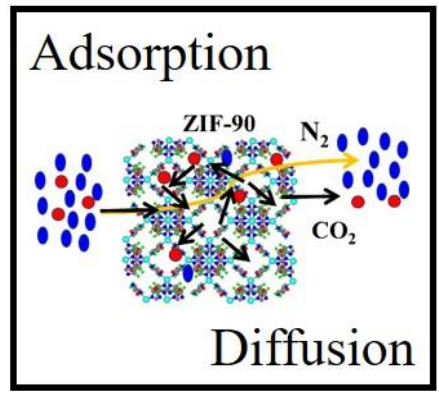

# 1. Introduction

Decarbonisation remains a key industrial commitment to reducing atmospheric $CO_2$ emissions. In this pursuit, adsorption-based processes are gaining prominence over conventional absorption due to their lower regeneration costs. Various porous materials, including zeolites, metal oxides, silica, carbon, metal-organic frameworks (MOFs), and organic polymers, have been investigated for $CO_2$ capture.[1] Among these, MOFs appear to be a versatile class of crystalline materials owing to their chemical modularity, high surface area, and tunable porosity, making them a promising class of material for gas separation and storage. Zeolitic imidazolate frameworks (ZIFs), which are a subfamily of MOFs with zeolite-like topologies,[2] possess exceptional thermal and chemical stability[3], and have been widely studied for $CO_2$ capture and separation from gas mixtures. In particular, ZIF-90, which is constructed from $Zn^{2+}$ metal nodes and imidazolate linkers with a carboxaldehyde functional group, exhibits a microporous architecture that has been explored for catalysis[4], gas separation[5], drug delivery[6], and sensing[7] applications. Unary adsorption studies of $CO_2$, $N_2$, and $CH_4$ in ZIF-90 highlight the critical role of polar aldehyde groups in enhancing $CO_2$ affinity, while the narrow pore apertures modulate adsorption energetics and molecular discrimination.[7,8] Such single-component adsorption measurements provide a useful first assessment for identifying materials with preferential $CO_2$ uptake. However, real flue gas streams contain multiple components, where competitive adsorption dominates and separation becomes far more complex. Mixture adsorption studies are therefore essential to understand the actual separation behaviour.

Functionalization of ZIF-90 has been shown to enhance $CO_2$ sorption capacity,[9–13] while co-adsorbed species such as $H_2O$ and $SO_2$ may introduce additional adsorption sites for $CO_2$, thereby improving selectivity over $N_2$[14]. Incorporating ZIF-90 in mixed matrix membranes has shown higher selectivity for $CO_2$ than $CH_4$ and $N_2$.[15,16] In practical post-combustion capture scenarios, the competition between $CO_2$ and $N_2$ is the key factor governing separation efficiency. Yet detailed molecular-level insights into their competitive behaviour within ZIF-90 remain limited. Molecular simulations can probe these interactions and understand adsorption-desorption processes in microporous frameworks. Prior molecular dynamics (MD) studies of adsorption in porous ZIF frameworks have largely focused on single-component systems[17–23], leaving the relationship between feed composition and adsorption composition insufficiently explored. Meanwhile, mixture studies broadly focus on the selectivity of a component from a mixture using GCMC simulations.[14,24–30] The diffusivities of all the commonly known gases - $CO_2$, $N_2$, $O_2$, $H_2$, CO, NO, NOx, $SF_6$, $SO_2$, $CH_4$, $C_2H_6$, $C_2H_4$, $C_3H_8$, $C_3H_6$, $C_4H_8$, $C_4H_{10}$, $C_6H_6$, $CH_3OH$, $CH_3CH_2OH$, $H_2$-$CH_4$, $CH_4$-$H_2$, $CO_2$-$CH_4$, $CO_2$-$N_2$, $CO_2$-$SO_2$, $CH_4$-$SO_2$, $CH_4$-$CO_2$, $O_2$-$C_2H_6$, $C_2H_4$-$C_2H_6$, $CH_3CH_2OH$-$H_2O$ in MOFs have been reported.[19,21,25,27,30–37] However, the diffusion behaviour of gas mixtures in ZIFs and, more importantly, the interdependence between the spatial distribution and diffusion of individual gas species within the ZIF-90 framework remains poorly understood.

Motivated by these questions, here, our objectives are to develop a systematic understanding of gas adsorption from a feed of $CO_2$ and $N_2$ mixtures, including their spatial distribution, diffusion and their interrelationship within the ZIF-90 porous framework. In our recent work,[8] we have developed integrated GCMC and MD simulation frameworks to investigate the adsorption and transport behavior of single-component gases within ZIF-90. In the present study, we extend this framework to multicomponent gas mixtures, enabling the characterization of competitive adsorption and diffusion phenomena that emerge under mixed-gas conditions. Specifically, we consider $CO_2$-$N_2$ binary mixtures with varying feed compositions, including flue gas relevant to practical conditions. Our GCMC simulations provide a quantitative correlation between feed composition and adsorbed composition of the gas mixture, while our MD simulations elucidate the interplay between selective adsorption and diffusion. Through systematic analysis of isotherms, selectivity, diffusion coefficients, and spatial distributions, we reveal how ZIF-90 balances strong $CO_2$ uptake with the transport limitations that govern regenerability. We also establish a connection between gas mixture diffusion between the bulk environment and the porous medium of ZIF-90. This study thus provides critical insights into the dual aspects of adsorption and desorption, offering a rational basis for evaluating ZIF-90 as a candidate material for energy-efficient $CO_2$ separation. Our results provide molecular-level insights into how competitive adsorption modulates gas distribution and transport inside the porous framework, offering a foundation for designing ZIF-90 and related materials for efficient gas separation applications.

## 2. Model and Methodology

We have obtained the unit cell of ZIF-90 from our previous study, wherein the structure is optimized using density functional theory (DFT) based electronic structure calculation.[8] The unit cell is replicated in all three directions to create a supercell of dimension 2*2*2. The total number of atoms in this supercell is 2016. We consider this supercell as the simulation box with periodic boundary conditions applied to all three directions. A systematic simulation workflow employed in this study, including the system consideration, adsorption equilibration, production, and diffusion evaluation, is shown in Figure 1. The universal force field (UFF) is used to model the interaction between the atoms of ZIF-90, which has been shown previously as a reasonably good representation of ZIF-90 framework[38]. $CO_2$ and $N_2$ are modelled using the EPM2 force field[39] and the TraPPE potential[40], respectively. The Lorentz-Berthelot mixing rule is used for cross interactions between gas and ZIF-90. The Lennard-Jones (LJ) and coulombic interactions are truncated at a cut-off distance of 13 Å. We used the Ewald summation method for the long-range electrostatic interaction calculations. A simulation involves two components - GCMC and MD, which are performed sequentially. The details of these two are provided below.

## 2.1. Grand Canonical Monte Carlo (GCMC) simulation

Binary mixture adsorption capacity is obtained by performing GCMC simulations using DL_MONTE open-source software (version. 2.0(7)/January 2020)[41]. The GCMC simulations are performed for a wide range of pressures, for a fixed temperature of 298 K. We perform $10^7$ MC steps for equilibrating the system, wherein sixty percent of MC moves are insertion/deletion, and thirty percent are translational, and the remaining are rotational moves. The imposed partial pressures of adsorbates were related to the Boltzmann distribution, and equilibrium was attained by applying the Metropolis criterion during configurational sampling. Following this equilibration, we perform a production run of $10^7$ MC steps. The production simulation data are used to obtain the adsorption capacity of the binary $CO_2/N_2$ mixture of various feed ratios in the ZIF-90. Further, the selectivities are estimated from uptakes as $Selectivity\ of\ \left(\frac{CO_2}{N_2}\right) = \frac{y_{CO_2}/y_{N_2}}{x_{CO_2}/x_{N_2}}$ where y and x are adsorbed and feed mole fractions of gas, while the total heat of adsorption is computed using the Green-Kubo fluctuation theory expression[42], $q = \frac{\langle UN \rangle - \langle U \rangle \langle N \rangle}{\langle N^2 \rangle - (\langle N \rangle)^2} + RT$ where $N$ is the adsorbed molecules, and $U$ is the total energy of a configuration.

## 2.2. Molecular Dynamics (MD) Simulations

The GCMC equilibrated $CO_2$-$N_2$ binary mixture in ZIF-90 configurations is used to study the adsorbed species' dynamic behaviour and diffusion properties within the ZIF-90. MD simulations are carried out using the Large-scale Atomic/Molecular Massively Parallel

Simulator (LAMMPS) package (June 23, 2022 and Update-2)[43]. An integration timestep of 1 fs is used for the MD simulation. The system is equilibrated for 10 ns in a canonical ensemble (NVT), followed by a production run of 10 ns in a microcanonical ensemble (NVE). The temperature is maintained at 298 K during the equilibration run using the Nose-Hoover thermostat with a damping factor of 100 fs. During the production run, all atom coordinates are collected every 1ps for post-analysis. We calculate the mean-squared displacement (MSD) of gases from the production trajectory as a function of time, which is used to compute the self-diffusion coefficients of gases (*cf.* Figure 1). We further compute the structure correlations, such as the radial distribution function (RDF) of sorbate molecules within the ZIF-90 framework that governs their diffusion in a porous material.

In addition, to provide a reference, we also perform MD simulations of bulk gas mixtures with compositions matching those of the confined systems. The ratio of $CO_2$ and $N_2$ molecules is thus maintained identical to that adsorbed within the ZIF-90 framework at different feed compositions under 1 bar pressure. Each bulk system undergoes a 10 ns equilibration simulation, followed by a 10 ns production run.

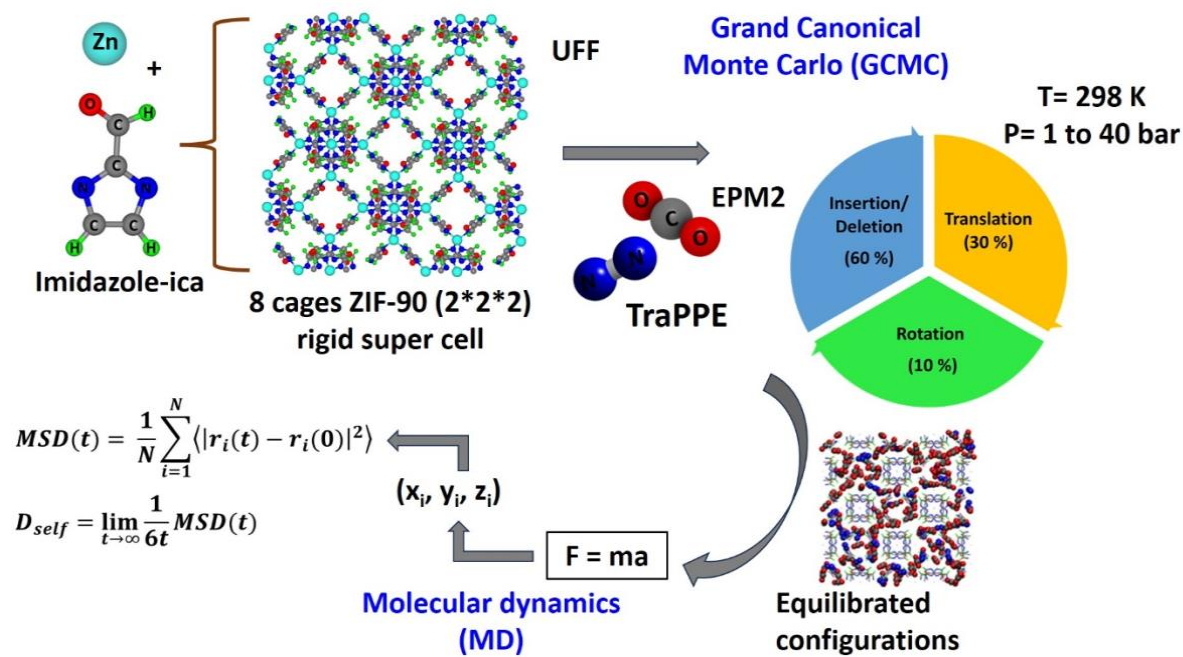

*Figure 1. Simulation workflow. The simulation box consists of 8 cages of ZIF-90, whose constituents are Zn and Imidazole Carboxaldehyde. A GCMC simulation is performed to compute the gas composition inside the simulation box for a given T and P. The GCMC-equilibrated structure served as the initial configuration for the subsequent MD Simulation in an NVT ensemble. The cage atoms are frozen throughout the simulation. MSD and Self-diffusion coefficients are computed from the MD trajectory.*

# 3. Results and Discussion

## 3.1. Gas adsorption and selectivity of the ZIF-90 framework

Our simulations of binary mixture adsorption in ZIF-90 show that $CO_2$ consistently adsorbs more than $N_2$ across a pressure range of 40 bar, as depicted in Figure 2a. This is attributed to the favourable interactions between the polar iso-carboxaldehyde group of imidazole in ZIF-90 and the quadrupole moment of $CO_2$, which boosts adsorption capacity (see section 3.1). For a 15/85 $CO_2$/$N_2$ mixture at 298 K and 1 bar, our measured $CO_2$ uptake (0.41 mmol g$^{-1}$) slightly exceeds the literature value (0.28 mmol g$^{-1}$)[14], likely due to differences in interaction modelling. Compared to pure gases, mixture uptakes are lower because of competitive adsorption at the sites. As the $CO_2$ fraction in the feed increases, the $CO_2$ uptake in ZIF-90 also rises, showing that the adsorption sites are mainly occupied by $CO_2$ rather than $N_2$, which improves the separation potential. We observe that the ZIF-90 exhibits a strong preference for $CO_2$ compared to $N_2$, emphasising its effectiveness for $CO_2$ separation from flue gases. This selectivity is mainly due to the polar imidazolate linkers, which are functionalized with aldehyde groups that enhance quadrupole-dipole interactions with $CO_2$ molecules. In contrast, $N_2$ interacts weakly with ZIF-90, lowering adsorption affinity. The $CO_2$/$N_2$ selectivity remains nearly constant across different feed compositions at atmospheric pressure, indicating that framework-guest interactions, rather than gas concentrations, mainly influence $CO_2$ adsorption. As shown in Figure 2b, ZIF-90 consistently favours $CO_2$ adsorption even as the $CO_2$ content in the feed varies. This steady, composition-independent selectivity at low pressure highlights ZIF-90's effectiveness as an adsorbent for $CO_2$ capture in real-world scenarios. Additionally, $CO_2$ selectivity increases with higher feed ratios, rising from 16 at 1 bar to 25 at 40 bar, depending on the $CO_2$/$N_2$ feed ratio. The obtained selectivity values agree with earlier reported computational findings by Hu et al.[14] and Amrouche et al.[44], confirming the accuracy of calculations and separation performance. We note that ZIF-90 demonstrates superior selectivity compared to other ZIFs like ZIF-8, ZIF-68, and ZIF-69. However, at elevated pressures, ZIF-90 and ZIF-69 exhibit similar selectivity.[26,45] This suggests ZIF-90's potential effectiveness for dry flue gas separation at room temperature.

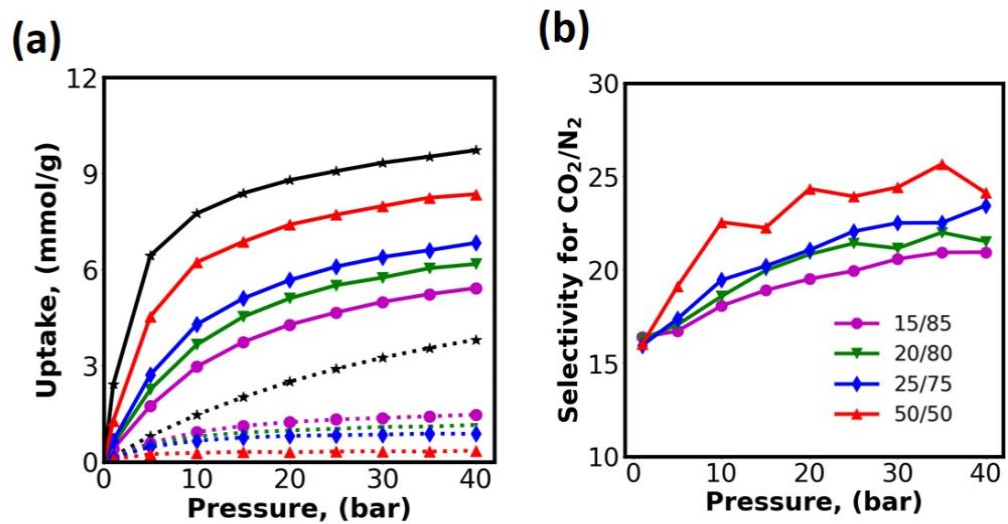

*Figure 2. (a) Isotherms of binary mixture ($CO_2/N_2$) adsorption at various feed ratios in ZIF-90. Solid lines and dashed lines represent carbon dioxide and nitrogen uptakes, respectively. Black solid and dashed lines represent pure $CO_2$ and $N_2$ uptakes. (b) Selectivity for $CO_2$ over $N_2$ in ZIF-90 at 298 K.*

The total heat of adsorption for the $CO_2/N_2$ mixture in ZIF-90 increases with uptake, following a trend similar to pure $CO_2$ adsorption (Figure 3a). In contrast, pure $N_2$ exhibits only a slight increase in the heat of adsorption with uptake, and its low heat of adsorption confirms weaker interactions with the framework. At lower uptakes, the heat of adsorption curves for mixtures with varying compositions exhibit slopes similar to that of pure $CO_2$, and these curves progressively shift toward the pure $CO_2$ curve as the $CO_2$ content in the feed increases. However, when the mixture uptake is ~ 7mmol/g, the heat of adsorption curves exhibit a steeper slope, nearly reaching the value of pure $CO_2$ adsorption, indicating that $CO_2$ contributions predominantly govern the total heat of adsorption. An inherent trade-off exists between selectivity and regenerability: strong $CO_2$-framework interactions improve selectivity and uptake but also result in higher heats of adsorption, thereby increasing the energy required for regeneration during PSA/TSA cycles. ZIF-90 shows better selectivity for $CO_2$ with a heat of adsorption below 30 kJ/mol, indicating good regeneration ability as an adsorbent for selective $CO_2$ capture.

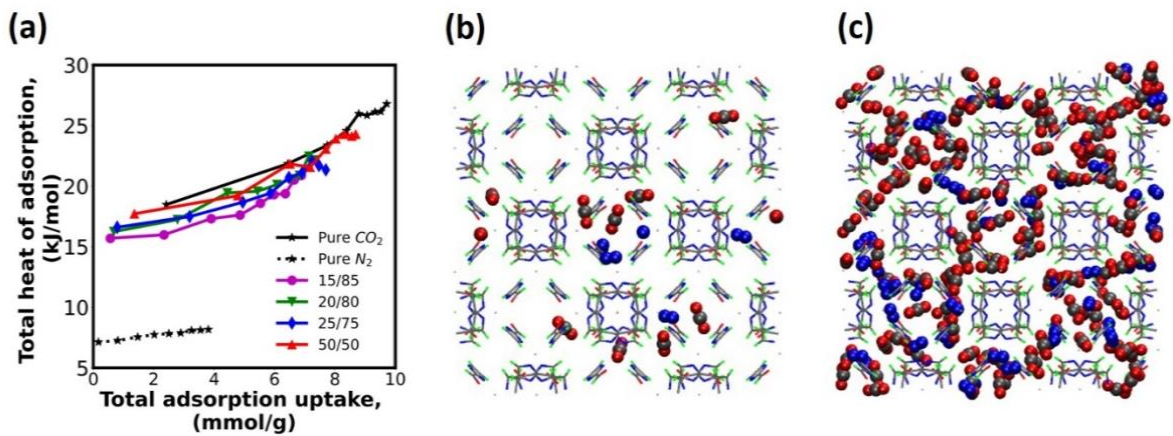

*Figure 3. (a) Total heat of adsorption of binary CO₂/N₂ mixture in ZIF-90. (b), (c) are GCMC equilibrated configurations of binary gas (CO₂/N₂) in ZIF-90 at 1 and 40 bar. (Grey, red, and blue coloured balls represent carbon, oxygen and nitrogen, while the ZIF-90 is shown in bond representation, highlighting framework topology.)*

## 3.2 Gas adsorption sites within the ZIF-90 framework

During binary adsorption, each adsorption site experiences competitive occupation by both sorbate species. Therefore, a thorough understanding of the underlying framework-gas and gas-gas interactions that govern the spatial distribution of the sorbates is crucial for the rational design of adsorbent materials. Such insights enable the development of frameworks capable of achieving high $CO_2$ selectivity while maintaining rapid regeneration performance. Figures 4(a-i) represent the RDFs of the $CO_2$ and $N_2$ gases from different atomic sites of the ZIF-90 framework at 1 bar and 298 K. Since we have two building units, such as metal (Zn) and organic ligand (Imidazolate) and functionalization as -CHO polar group in ZIF-90, we considered these as the references to assess the relative positions of both carbon and oxygen atoms of $CO_2$ and centre of mass of the $N_2$ molecule to determine the spatial arrangement of gas molecules within the eight cages of ZIF-90.

The imidazole linker's polar -CHO functional group exhibits a stronger affinity toward the carbon atom of $CO_2$, as evidenced by a prominent interaction peak at approximately 3.5 Å with an intensity greater than 2 (Figure 4d). In contrast, the interaction with the oxygen atom of $CO_2$ occurs at a larger distance of around 4.0 Å and with a weaker intensity (< 2) (Figure 4d). The GCMC equilibrated configurations confirm that the $CO_2$ was more adsorbed at the organic ligands of ZIF-90, as shown in Figures 3b and 3c. Also, the double-bounded carbons offer favourable interactions with the oxygens of $CO_2$, confirmed by the first coordination at 3.8 Å (Figure 4b), than with $N_2$ at a distance of 4.2 Å (Figure 4c). Additionally, the interaction peaks corresponding to $CO_2$, particularly those involving the double-bonded carbon, functionality carbon and zinc, appear at different positions in ZIF-90 (3.6 Å, 3.8 Å, and 6.2 Å) compared to ZIF-8 (approximately 4.4 Å, 4.0 Å, and 6.2 Å, respectively).[25] This variation arises despite both frameworks sharing similar structural constituents- zinc nodes coordinated by imidazolate linkers. The difference originates from the nature of the functional groups attached to the imidazolate linker: the polar -CHO group in ZIF-90 enhances the electrostatic interactions with the quadrupolar $CO_2$ molecules through dipole-quadrupole attraction, leading to stronger binding. In contrast, the nonpolar $-CH_3$ group in ZIF-8 lacks such polarity and therefore exhibits weaker interactions with $CO_2$. Consequently, ZIF-90 provides more favourable adsorption sites and higher affinity for $CO_2$, reflecting the critical role of linker functionalization in tuning gas-framework interactions. Overall, $CO_2$ has a favourable adsorption through imidazole's polar -CHO functionality compared to metal sites in ZIF-90. These interactions explain the moderate heat of adsorption of $CO_2$ in ZIF-90.

The $N_2$ adsorption heat is low due to weaker interactions with organic ligands, as shown by the RDF at 4 Å with an intensity below 2, depicted in Figure 4c. In contrast, $CO_2$ (Figures 4d and 4e) displays higher intensities exceeding 2. $N_2$ also favours metal sites (Figure 4i), along with $CO_2$ (Figures 4g and 4h). This supports our hypothesis that increasing $CO_2$ in the feed will replace weakly adsorbed $N_2$ sites with $CO_2$ (see Figure 2(b) at high pressures), resulting in higher uptakes and selectivity. Since each Zn metal is coordinated with four organic ligands (Ica-Imidazole), they provide more adsorption sites than the metals. Multiple peaks in the RDF plots suggest coordination akin to the framework's secondary and tertiary adsorption sites.

The pair correlation functions between gas molecules and the ZIF-90 framework remain almost unchanged across different feed compositions. This indicates that the local structural environment around the adsorbed molecules remains largely unaffected by changes in gas mixture ratios. Essentially, ZIF-90's adsorption sites preserve a consistent affinity for $CO_2$ regardless of the $CO_2/N_2$ ratio in the feed. This behaviour stems from strong, localised interactions between $CO_2$ and the polar -CHO groups in ZIF-90, along with potential gas-gas interactions (see section 3.2). The invariance in the pair correlation profiles indicates that the framework provides well-defined, energetically accessible adsorption sites structurally resistant to competition. Consequently, $CO_2$ binds preferentially across various compositions, supporting the selectivity observed in the isotherms and mixture adsorption simulations.

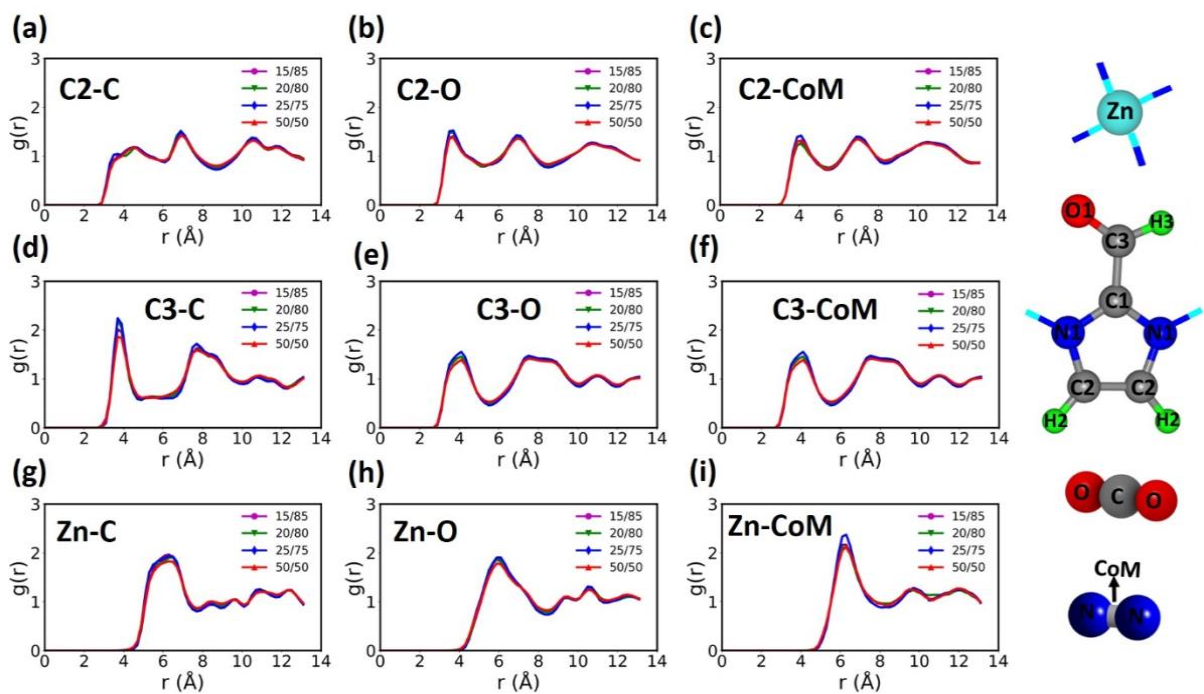

*Figure 4. Structural ordering of $CO_2$ and $N_2$ with respect to different atomic sites of the ZIF-90 framework at 1 bar and 298 K for (15/85) feed ratio. The atomic sites are the metal centre and the functional group on the linkers. (a), (d), and (g) are based on carbon atom in the $CO_2$ molecule, (b), (e), and (h) are based on oxygen in the $CO_2$ molecule, while (c), (f), and (i) are based on centre of mass of nitrogen molecule with respect to double-bonded carbons, iso-carboxaldehyde carbon, and zinc in ZIF-90.*

## 3.3 Gas-gas interaction and their spatial correlations

In gas mixture adsorption systems, guest molecule behaviour is influenced by interactions with the framework and significantly affected by gas-gas interactions within the confined cage environment. These interactions determine how sorbate molecules rearrange spatially during competitive adsorption. We compute the pair correlation functions (RDFs), for $CO_2$-$CO_2$, $N_2$-$N_2$, and $CO_2$-$N_2$ pairs to provide quantitative insights into their relative arrangements, the level of association and interaction between them. These gas-gas correlations are crucial for their selectivity, uptake, and cooperative behaviour in mixed-gas adsorption, Figure 5 presents the gas-gas pair correlation functions for the bulk gas mixture and within the ZIF-90 framework, calculated up to 13 Å. RDF analysis shows lateral adsorbate-adsorbate interactions are more prominent inside ZIF-90 than in the bulk mixture. The $CO_2$-$CO_2$ correlations, with peaks at 4 Å and 9 Å (Figure 5a), suggest a tendency for $CO_2$ molecules to cluster, influenced by the framework's polar aldehyde groups. The $N_2$-$N_2$ correlations remain strong across different feed ratios (Figure 5c), indicating limited interactions with ZIF-90 compared to $CO_2$-$CO_2$ (Figure 5a). Additionally, in ZIF-90, $CO_2$-$N_2$ interactions are less intense at low feed ratios but become stronger as the local $CO_2$ concentration increases with higher feed ratios. (Figure 5b). These apparent coordination effects were not observed in the bulk gas mixture dynamics, confirming the confinement effects of the framework ZIF-90. The enhanced $CO_2/N_2$ selectivity with the feed ratio was intuitive as $CO_2$ molecules experience framework affinity and cooperative lateral interactions. However, these same strong interactions can hinder molecular mobility within ZIF framework cages, which in turn slows desorption and may prolong regeneration cycles. Together, these insights establish a clear link between the microscopic structuring of adsorbates in ZIF-90 and the thermodynamic selectivity toward $CO_2$ during mixture adsorption.

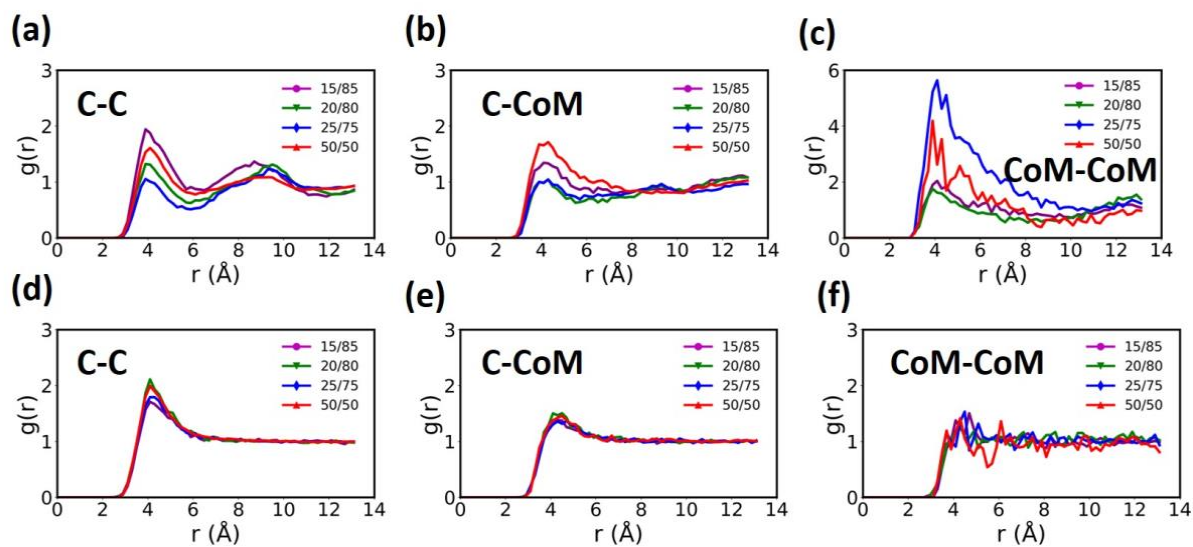

*Figure 5. Gas-gas correlation was represented by the radial distribution functions where (a) and (d) are based on carbon-to-carbon atoms in the $CO_2$ molecules, (b) and (e) are based on carbon atom in $CO_2$ with the centre of mass of $N_2$, while (c) and (f) are based on the centre of mass to the centre of mass in the $N_2$ molecules. (a-c) and (d-f) corresponds to a binary mixture adsorbed in the ZIF-90 framework and bulk gas without the framework at 1 bar and 298 K.*

## 3.4 Gas transport behaviour within the ZIF-90 framework

The mean square displacement (MSD) analysis derived from MD simulations offers important insights into the diffusion behaviour of $CO_2$ and $N_2$ in ZIF-90 under competitive adsorption conditions. During the bulk phase, molecular motion is determined only by gas-gas interactions at constant thermodynamic conditions. In contrast, gas-gas and ZIF-gas interactions within the framework limit mobility. As a result, more robust adsorbate-framework interactions lead to slower molecular movement. The MSD profiles reveal that $CO_2$ molecules move less freely than $N_2$, implying they experience stronger confinement and interact more intensely with the framework. This also aligns with findings as $CO_2$ tends to occupy energetically favourable adsorption sites, where the weakly adsorbed $N_2$ restricts movement through site blocking and competitive adsorption. As the $CO_2$ concentration in the mixture increases, a further reduction in the diffusivity of both gases is observed, reflecting intensified intermolecular interactions and framework loading. Figures 6a and 6b illustrate that the $CO_2/N_2$ mixture adsorbed in ZIF-90 requires approximately $10^6$ ps to attain the diffusive regime, while the bulk mixture reaches this stage in about $10^4$ ps, as shown in Figures 6c and 6d, with the MSD growing linearly over time (slope ≈ 1). This analysis demonstrates that confinement within ZIF-90 decreases molecular displacement by nearly five orders of magnitude relative to the bulk phase.

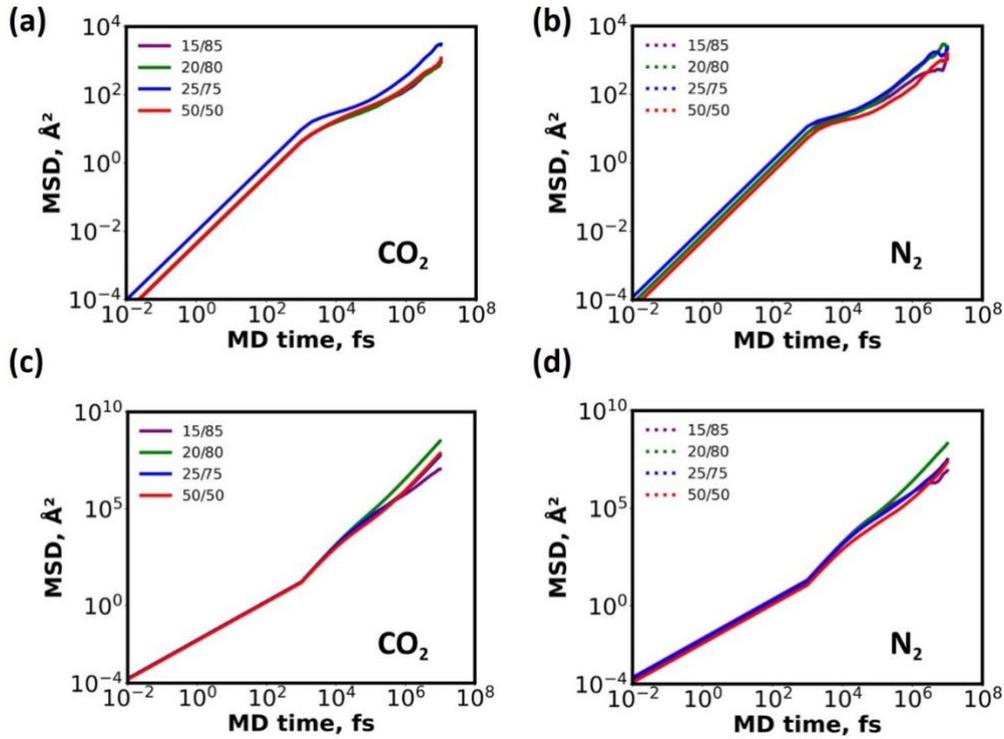

Figure 6. Mean square displacement of gases. (a) and (b) are adsorbed $CO_2$ and $N_2$, binary mixtures in the ZIF-90 framework, while (c) and (d) are $CO_2$ and $N_2$ in bulk binary gas mixtures without framework at 1 bar and 298 K.

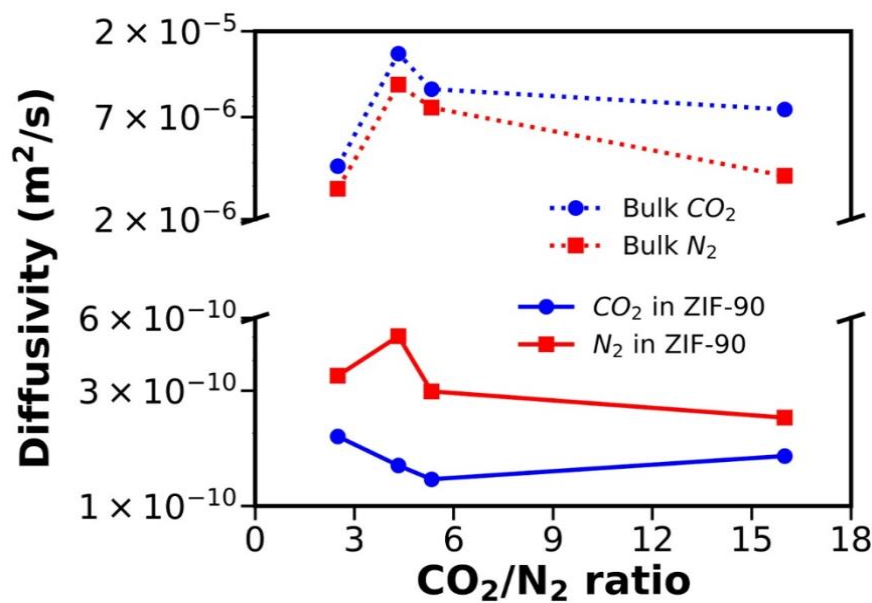

*Figure 7. Self-diffusivities of $CO_2$ and $N_2$ are plotted as a function of the mixture composition ($CO_2/N_2$ ratio) within the framework. The diffusivities of the gases in bulk mixture are also plotted for the identical composition. The error bars are smaller than the symbol sizes. All the data are at 1 bar and 298 K.*

In bulk gas mixtures, diffusion is mainly influenced by kinetic diameter and collision behaviour rather than molecular weight alone. Although $CO_2$ (44 g/mol) is heavier than $N_2$ (28 g/mol), it can have similar or even higher bulk diffusivity because of its smaller kinetic diameter (3.3 Å) and larger quadrupole moment (4.3 D Å).[46] This enhances its mobility in the gas phase. Yet, in ZIF-90, this trend reverses; cages with ligands containing polar aldehyde groups form strong host-guest interactions with $CO_2$ (see Figure 4d), which hinder its movement. $N_2$, with weaker interactions (< 10 kJ/mol, see Figure 3a), travels faster through the cages. As a result, Figure 7 clearly demonstrates the reversal of diffusion coefficients for $CO_2$ and $N_2$ between the bulk and the ZIF framework. Quantitatively, at 298 K and 1 bar, the pure gas diffusivity of $CO_2$ in ZIF-90 was reported as $1.83 \times 10^{-9}$ m$^2$/s, while in our (15/85) $CO_2/N_2$ mixture case, it was $1.94 \times 10^{-10}$ m$^2$/s. This analysis demonstrates how a co-adsorbate affects the system. It shows that while strong adsorption improves selective $CO_2$ capture, it also impedes transport, leading to longer regeneration times. These results confirm that MSD-based diffusivity estimation accurately reflects the tension between competitive adsorption and transport processes, offering detailed insight into the diffusion of binary mixtures in ZIF-90.

## 4. Conclusions

In this work, we have systematically investigated the adsorption and transport behaviour of $CO_2/N_2$ mixtures in ZIF-90 using a combination of GCMC and MD simulations. Mixture adsorption simulations revealed that $CO_2$ preferentially adsorbs over $N_2$ due to strong framework interactions, as evidenced by RDF analysis, while $N_2$ diffuses faster owing to weaker host-guest interactions. MSD calculations showed that confinement within ZIF-90 slows molecular diffusion by several orders of magnitude compared to bulk, indicating a diffusion-adsorption trade-off that impacts regeneration. Heat of adsorption analysis further confirmed that $CO_2$ governs the total adsorption energetics, underscoring the balance between selectivity and energy requirement during PSA/TSA regeneration. Overall, these results provide molecular-level insights into how ZIF-90 achieves selective $CO_2$ capture, highlighting its potential as a candidate material for energy-efficient post-combustion gas separation processes.

**Acknowledgements**

We acknowledge the use of computing resources of the HPCE, IIT Madras. TKP acknowledges SERB (ANRF) for support through a core research grant (CRG/2022/006926).